%% file: manuscript.tex
\documentclass[12pt]{iopart}

\usepackage{setstack}
\usepackage{amssymb}
\usepackage[]{graphicx}
\usepackage{mathrsfs}
\usepackage{bm}
\usepackage{braket}

\usepackage[]{cite}

\usepackage[T1]{fontenc}

\usepackage{cases}

\usepackage{comment}

\begin{document}

\title{Effects of surface plasmons on spin currents in a thin film system}

\author{Daigo Oue $^{1,2}$ \& Mamoru Matsuo $^{1,3,4,5}$}

\address{$^1$ Kavli Institute for Theoretical Sciences, University of Chinese Academy of Sciences, Beijing, 100190, China.}
\address{$^2$ The Blackett Laboratory, Department of Physics, Imperial College London, Prince Consort Road, Kensington, London SW7 2AZ, United Kingdom}
\address{$^3$ CAS Center for Excellence in Topological Quantum Computation, University of Chinese Academy of Sciences, Beijing 100190, China}
\address{$^4$ RIKEN Center for Emergent Matter Science (CEMS), Wako, Saitama 351-0198, Japan}
\address{$^5$ Advanced Science Research Center, Japan Atomic Energy Agency, Tokai, 319-1195, Japan.}
\ead{daigo.oue@gmail.com}
\vspace{10pt}
\begin{indented}
\item[] \today
\end{indented}

\begin{abstract}
 We propose and analyze surface-plasmon-driven electron spin currents in a thin metallic film. 
 The electron gas in the metal follows the transversally rotating electric fields of the surface plasmons (SPs),
 which leads to a static magnetization gradient. 
 We consider herein SPs in a thin-film insulator-metal-insulator structure and solve the spin diffusion equation in the presence of a magnetization gradient.
 The results reveal that the SPs at the metal interfaces generate spin currents in the metallic film.
 For thinner film,
 the SPs become strongly hybridized,
 which increases the magnetization gradient and enhances the spin current.
 We also discuss how the spin current depends on SP wavelength and the spin-diffusion length of the metal.
 The polarization of the spin current can be controlled by tuning the wavelength of the SPs and/or the spin diffusion length. 
\end{abstract}
\submitto{\NJP}
\maketitle

\section{Introduction}
Surface plasmons (SPs) are excitations localized at a metal-dielectric interface and are composed of electromagnetic waves coupled with plasma oscillations of the electron gas in the metal.
SPs propagate along the metal-dielectric interface as evanescent waves in both the metallic and dielectric parts of the interface.
Note that evanescent waves have transverse spin,
in which the electric field rotates in the direction perpendicular to the propagation direction \cite{bliokh2014extraordinary},
whereas the spin of an ordinary plane wave is parallel or antiparallel to its propagation direction.
The spin direction is uniquely determined by the propagation direction and the decay direction \cite{van2016universal}. 
This is the so-called spin momentum locking effect of evanescent waves,
and is observed in many systems,
such as in total internal reflection configuration,
in optical fibres,
and lossy interfaces \cite{antognozzi2016direct, kalhor2016universal, fang2017intrinsic, oue2019dissipation}.
SPs are no exception to this trend and have transverse spin in both the metal and the dielectric media \cite{bliokh2012transverse, rodriguez2013near, canaguier2014transverse, bliokh2015quantum, triolo2017spin, dai2018ultrafast}.

\par
Because the SP frequency is below the plasma frequency,
the electron gas coupled with a SP follows the circularing electric field of the SP,
and the orbiting motion of the electrons thus-generated produces a static and inhomogeneous magnetization in the metal because of the evanescent intensity profile of SPs.
Since the electric current generated by inhomogeneous magnetization has zero divergence,
this magnetization has been considered undetectable by electrical measurements\cite{bliokh2017optical_new_j_phys}.
However, we reveal that not only charge currents but also electron-spin currents are generated by this magnetization and that both can be detected.

\par
In general,
a metal can support two types of electronic transport: charge currents and spin currents.
Recently,
the spintronics community has reported that inhomogeneous effective magnetic fields generate spin currents\cite{kohda2012spin, takahashi2016spin, kobayashi2017spin, okano2019nonreciprocal}.
Electron spin currents are generated when electrons are driven by spin-dependent force potentials created by an inhomogeneous effective magnetic field (e.g., spin-orbit coupling \cite{kohda2012spin} and spin-vorticity coupling \cite{takahashi2016spin, kobayashi2017spin, okano2019nonreciprocal}).
In other words,
a Stern-Gerlach-like effect drives the electron-spin current.

\par
\begin{figure}[tbhp]
  \centering
  \includegraphics[width=10cm]{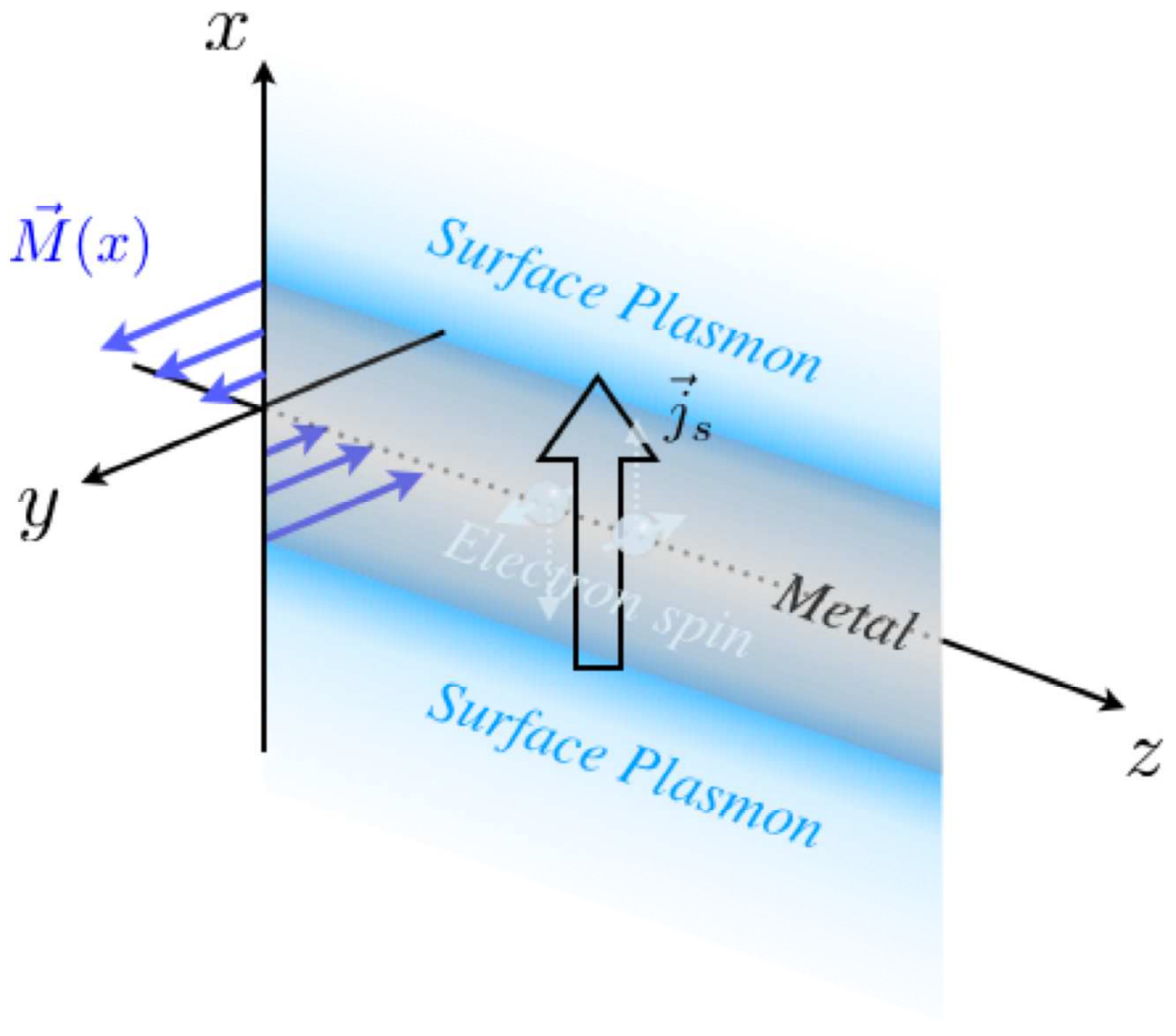}
  \caption{
    Angular-momentum conversion from surface plasmon to electron spin in a metallic thin-film system. 
    The film thickness is $d$.
    The upper surface is at $x=+d/2$ and the lower surface is at $x=-d/2$.
    In the metallic film,
    the surface plasmon mode on the upper surface is hybridized with that on the lower surface.
    The electric field of a hybridized surface plasmons rotates transversally in the metallic film,
    thereby creating at inhomogeneous magnetization $\vec{M}(x)$ via the inverse Faraday effect \cite{bliokh2017optical_new_j_phys, pitaevskii1961electric, nkoma1974elementary, kono1981spontaneous}.
    This inhomogeneous magnetization drives the electron spin current $\vec{j}_s$,
    which exerts a Stern-Gerlach-like effect on the conduction-electron spins in the metal.
  }
  \label{fig:am_conversion}
\end{figure}

In this paper,
we consider SPs in a thin film of non-magnetic metal and investigate whether the SPs drive spin currents (see FIG. \ref{fig:am_conversion}).
There is a mechanism where plasmon-magnon interaction produces spin currents at metal-magnetic material interfaces \cite{uchida2015generation, ishii2017wavelength}.
Compared to these previous studies, 
the significant point in our study is that the transverse spin of the SPs in the thin film together with the spin-momentum locking can be an alternative way to generate pure spin currents from light without any magnetic field or magnetic substances.
We solve the spin diffusion equation with inhomogeneous magnetization created by the SPs and show that SPs in a thin film lead to spin accumulation and diffusive currents.
We use Gaussian units throughout this paper.

\section{Surface plasmons in a thin film}
Following the literature \cite{maier2007plasmonics}, we derive the dispersion relations and the corresponding eigenmodes in order to calculate the spin angular momenta of the eigenmodes in the next section.
We begin by considering a transverse-magnetic mode propagating in the $+z$ direction in a thin film system:
\begin{equation}
  \vec{E} = \left(
  \begin{array}{c}
    E_x(x)\\
    0\\
    E_y(x)
  \end{array}\right) e^{ikz}, \quad 
  \vec{H} = \left(
  \begin{array}{c}
    0\\
    H_y(x)\\
    0
  \end{array}\right) e^{ikz},
\end{equation}
where we have defined the wavenumber $k$.
Note that we omit the time-dependent factor $e^{-i\omega t}$ throughout this paper.
Using the monochromatic Maxwell equations
\begin{subnumcases}
  {}
  \nabla \cdot \vec{E} = \nabla \cdot \vec{H} = 0,\\
  \nabla \times \vec{H} = -i\frac{\omega}{c}\epsilon \vec{E},\\
  \nabla \times \vec{E} = i\frac{\omega}{c}\mu \vec{H},
\end{subnumcases}
gives simultaneous equations for the field amplitudes: 
\begin{subnumcases}
  {}
  E_z(x) = \frac{i}{k} \frac{\partial}{\partial x} E_x(x), \label{eq:Ez-Ex}\\
  H_y(x) = \frac{\omega}{ck} \epsilon E_x(x), \label{eq:Hy-Ex}\\
  \frac{\partial^2}{\partial x^2} E_x(x) = K^2 E_x(x). \label{eq:Ex_wave_eq}
\end{subnumcases}
Here, we have defined $K \equiv \left(k^2 - \omega^2 \epsilon / c^2\right)^{1/2}$.
The solutions to Eq. (\ref{eq:Ex_wave_eq}) in each region is
\begin{eqnarray}
E_x(x)=\left\{ \begin{array}{ll}
A_- e^{K_i (x+d/2)} & x < -d/2,\\
A_m' e^{-K_m (x+d/2)} + A_m'' e^{K_m (x-d/2)} & |x| < d/2,\\
A_+ e^{-K_i (x-d/2)} & d/2 < x.
\end{array} \right.
\label{eq:Ex}
\end{eqnarray}
We use
\begin{equation}
  K_{i,m} = \sqrt{k^2 - \frac{\omega^2}{c^2} \epsilon_{i,m}} = K_{i,m} (k, \omega)
\end{equation}
where $\epsilon_i$ and $\epsilon_m$ are the permittivities of the insulator and the metal, respectively.
Because we are interested in waves that are localized at the film surface,
we choose solutions that vanish as $x \rightarrow \pm \infty$.
Calculating the transverse-field quantities, $E_z(x)$ and $H_y(x)$, from Eqs. (\ref{eq:Ez-Ex}), (\ref{eq:Hy-Ex}), and (\ref{eq:Ex}) and imposing the standard continuity conditions of the transverse fields at $x=\pm d/2$,
we can get the following simultaneous equations in the matrix form:
\begin{eqnarray}
  \left(
  \begin{array}{cccc}
    K_i & K_m & -K_m e^{-K_m d} & 0\\
    0 & -K_m e^{-K_m d} & K_m & K_i\\
    \epsilon_i & -\epsilon_m & -\epsilon_m e^{-K_m d} & 0\\
    0 & \epsilon_m e^{-K_m d} & \epsilon_m & -\epsilon_i
  \end{array}
  \right)
  \left(
  \begin{array}{c}
    A_-\\
    A_m'\\
    A_m''\\
    A_+
  \end{array}
  \right)
  = 0
  \label{eq:eigenval_eq}
\end{eqnarray}
The condition for the existence of nontrivial solutions to Eq. (\ref{eq:eigenval_eq}) is
\begin{equation}
  \det \left(
  \begin{array}{cccc}
    K_i & K_m & -K_m e^{-K_m d} & 0\\
    0 & -K_m e^{-K_m d} & K_m & K_i\\
    \epsilon_i & -\epsilon_m & -\epsilon_m e^{-K_m d} & 0\\
    0 & \epsilon_m e^{-K_m d} & \epsilon_m & -\epsilon_i
  \end{array}
  \right) = 0,
\end{equation}
which gives the dispersion relation for SPs in the metallic-film system:
\begin{equation}
  \frac{1+R}{1-R} = \pm e^{-K_m d}. \label{eq:dispersion}
\end{equation}
In (\ref{eq:dispersion}), $R \equiv (K_m/\epsilon_m)/(K_i/\epsilon_i)$.
The corresponding eigenmodes are
\begin{equation}
\left(
   \begin{array}{c}
    A_-\\
    A_m'\\
    A_m''\\
    A_+
  \end{array}
  \right) 
=
\left(
  \begin{array}{c}
    -1\\
    -\frac{R-1}{2}\ \frac{K_i}{K_m}\\
    \frac{R-1}{2}\ \frac{K_i}{K_m}\\
    1
  \end{array}\right),
  \left(
  \begin{array}{c}
    1\\
    \frac{R-1}{2}\ \frac{K_i}{K_m}\\
    \frac{R-1}{2}\ \frac{K_i}{K_m}\\
    1
  \end{array}\right).
  \label{eq:eigenmodes}
\end{equation}
The first eigenmode in Eq. (\ref{eq:eigenmodes}) gives the dispersion relation $(1+R)/(1-R) = +e^{-K_m d}$,
and the second one gives $(1+R)/(1-R) = -e^{-K_m d}$.

\par
\begin{figure}[tbhp]
  \centering
  \includegraphics[width=.9\linewidth]{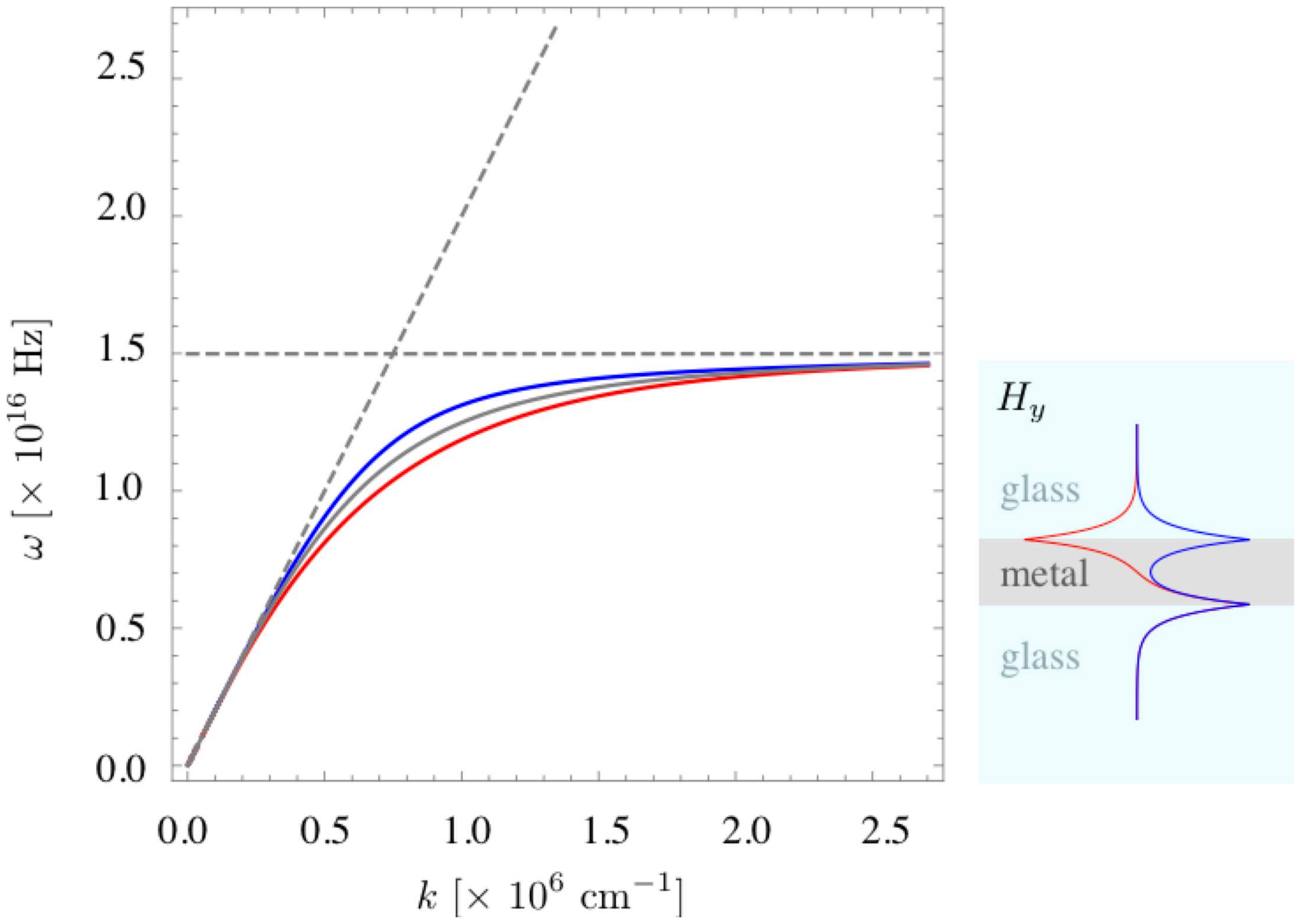}
  \caption{
    Dispersion relation of surface plasmons (SPs) in a thin metallic film.
    SPs are on the upper surface and on the lower surface;
    they are degenerate for infinitely thick films,
    but hybridize and split into two branches for thinner films.
    The upper branch (blue) corresponds to $(1+R)/(1-R) = +e^{-K_m d}$,
    and the lower branch (red) corresponds to $(1+R)/(1-R) = -e^{-K_m d}$.
    The right figure shows the magnetic-field distributions $H_y (x)$ for each mode at $k = 3\omega_p/c$.
    The field distribution is symmetric on the upper branch (red) and antisymmetric on the lower branch (blue).
    The solid gray curve is the dispersion relation for SPs on a single surface (without hybridization).
    The light line ($\omega = ck/\sqrt{\epsilon_i}$) and the surface plasma frequency in glass ($\omega = \omega_\mathrm{sp} \equiv \omega_p/\sqrt{1+\epsilon_i}$) are also shown.
    This plot uses the Drude parameter for gold ($\omega_p = 2.1 \times 10^{15}\ \mathrm{Hz}$),
    the permittivity for glass ($\epsilon_i = 2.25$),
    and a film thickness of $d = 20\ \rm{nm}$.
    }
  \label{fig:dispersion_SP}
\end{figure}

FIG. \ref{fig:dispersion_SP} shows the dispersion curves and the corresponding field distributions of SPs in a metallic-film system.
The dispersion curve has an upper branch $\omega_+(k)$ and a lower branch $\omega_-(k)$,
which we call the antibinding mode and the binding mode, respectively.
This splitting is due to the hybridization between the SP on the upper interface and the SP on the lower interface,
both of which are originally subject to the same dispersion (see the gray curve in FIG. \ref{fig:dispersion_SP}).
The splitting increases as the thickness of the film thins because the SPs are hybridized more strongly.
As depicted in FIG. \ref{fig:dispersion_SP},
the field distribution is symmetric (antisymmetric) on the antibinding (binding) branch,
which implies that the distribution of electric charge in the film is symmetric (antisymmetric) on the antibinding (binding) branch.
Depending on the mode,
the electrostatic interaction between the two interfaces is either repulsive or attractive
(which explains why the mode on the upper branch is called the antibinding mode,
and the mode on the lower branch is called the binding mode).

\section{Inhomogeneous magnetization induced by surface plasmon}
We now investigate the spin-angular-momentum (SAM) density of SPs on a thin metallic film.
We use the Minkowski representation to calculate the SAM:
\begin{equation}
  \vec{S} = \frac{g}{2} \mathfrak{Im} \left( \tilde{\epsilon} \vec{E}^* \times \vec{E} + \tilde{\mu} \vec{H}^* \times \vec{H} \right),
\end{equation}
where we use a Gaussian-unit factor  $g = (8 \pi \omega)^{-1}$,
group permittivity $\tilde{\epsilon} = \frac{\mathrm{d} (\omega \epsilon)}{\mathrm{d}\omega}$,
and permeability $\tilde{\mu} = \frac{\mathrm{d} (\omega \mu)}{\mathrm{d}\omega}$.
The use of the group permittivity and permeability corrects the dispersion of the SAM density.

\par
As previously shown in the literature \cite{bliokh2017optical_new_j_phys},
we can decompose the SAM density of the SP into two contributions:
one from the electromagnetic field and one from the kinetic motion of the electron gas:
\begin{eqnarray}
  \vec{S} = \vec{S}_{em} + \vec{S}_{mat},\\
          = \frac{g \epsilon}{2} \mathfrak{Im} \left( \vec{E}^* \times \vec{E} \right) + \frac{g\omega}{2} \frac{\mathrm{d} \epsilon}{\mathrm{d} \omega} \mathfrak{Im} \left( \vec{E}^* \times \vec{E} \right).
\end{eqnarray}
Note that the magnetic field of the SP does not contribute to the SAM because it does not rotate but just oscillates linearly.
The electron gas in the film undergoes kinetic motion and possesses angular momentum,
which magnetizes the film.
To determine the magnetization,
we multiply the electron contribution to the SAM by the gyromagnetic ratio \cite{herzberg1944atomic}:
\begin{equation}
  \vec{M} = \frac{-e}{2mc} \vec{S}_{mat} = -\frac{ge\omega}{4mc} \frac{\mathrm{d} \epsilon}{\mathrm{d} \omega} \mathfrak{Im} \left( \vec{E}^* \times \vec{E} \right).
  \label{eq:magnetisation}
\end{equation}

\par
From Eqs. (\ref{eq:eigenmodes}), (\ref{eq:Ex}), and (\ref{eq:Ez-Ex}),
we obtain the expressions for the electric fields in the metal $(|x| < d/2)$.
The antibinding mode and the binding mode are respectively
\begin{equation}
  \vec{E}_+ = (R-1) 
  \left(
  \begin{array}{c}
    \frac{K_i}{K_m}k \sinh (K_m x) \\
    0\\
    iK_i \cosh (K_m x)
  \end{array} \right)\Phi_0 e^{ikz-\frac{K_m d}{2}},
  \label{eq:E_film_antibinding}
\end{equation}
and
\begin{equation}
  \vec{E}_- = (R-1) 
  \left(
  \begin{array}{c}
    \frac{K_i}{K_m}k \cosh (K_m x) \\
    0\\
    iK_i \sinh (K_m x)
  \end{array}\right) \Phi_0 e^{ikz-\frac{K_m d}{2}}.
  \label{eq:E_film_binding}
\end{equation}
Note that we have regularized the electric-field vectors by multiplying them by $k\Phi_0$,
where $\Phi_0$ is the strength of the electric field.
We use Eqs. (\ref{eq:E_film_antibinding}) and (\ref{eq:E_film_binding}) to evaluate $\mathfrak{Im} \left( \vec{E}^* \times \vec{E} \right)$ and find
\begin{equation}
  \mathfrak{Im} \left( \vec{E}_{\pm}^* \times \vec{E}_{\pm} \right)
  =  
  \frac{k (K_i^\pm)^2}{K_m^\pm}\ |R^\pm-1|^2 |\Phi_0|^2 e^{-K_m^\pm d} \sinh (2K_m^\pm x) \vec{u}_y.
  \label{eq:Im}
\end{equation}
where the superscripts $^\pm$ on the right-hand side $^\pm$ represents the antibinding mode and the binding mode [e.g.$K_i^\pm = K_i(k,\omega_\pm)$], respectively.
The vector $\vec{u}_y$ is the unit vector in the $+y$ direction.
The fact that both modes have the same SAM expression stems from the spin-momentum locking of evanescent waves localized near interfaces \cite{van2016universal}.
For both the antibinding mode and the binding mode,
the decay direction of the wave on the lower interface is opposite that on the upper interface,
whereas the waves on the lower and upper interfaes propagate in the same direction.
The spin direction of a localized electric field can thus be uniquely determined by the decay direction and the propagation direction.

\par
To proceed to calculate the magnetization,
we use the Drude free-electron model for the permittivity of the metal:
\begin{equation}
  \epsilon = 1 - \frac{\omega_p^2}{\omega^2}.
  \label{eq:Drude}
\end{equation}
From Eqs. (\ref{eq:magnetisation}), (\ref{eq:Im}), and (\ref{eq:Drude}),
we can obtain the magnetization induced by a SP in a metallic film:
\begin{eqnarray}
  \vec{M}_\pm
  = 
  -\frac{g e \omega_p^2}{2 m c \omega_\pm^2}\ \frac{k (K_i^\pm)^2}{K_m^\pm}\ |R^\pm-1|^2 |\Phi_0|^2 e^{-K_m^\pm d} \sinh (2K_m^\pm x) \vec{u}_y,\\
  = M_0 f(k,d) \sinh \left( 2 K_m^\pm x \right) \vec{u}_y,
  \label{eq:magnetisation_thin_film}
\end{eqnarray}
where we used $M_0 = -(e|\Phi_0|^2)/(2mc)$ and $f(k,d) = g k \{|R^\pm-1|\ \omega_p K_i^\pm\}^2 e^{-K_m^\pm d} / (\omega_\pm^2 K_m^\pm)$.
Note that a larger magnetization is generated in the thinner film because of the exponential term $e^{-K_m^\pm d}$.
\begin{figure}[tbhp]
  \centering
  \includegraphics[width=.9\linewidth]{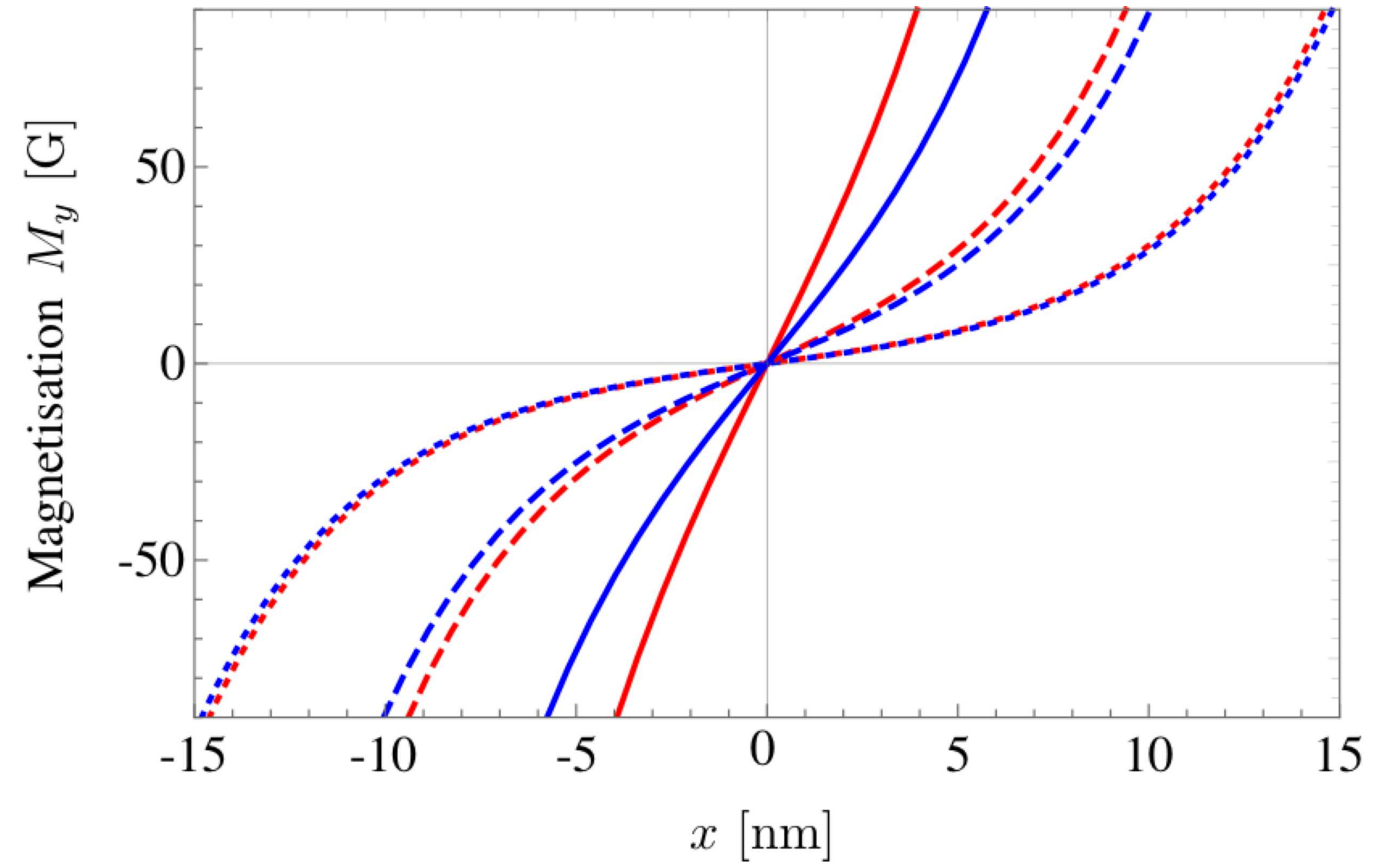}
  \caption{
    Distribution of SP-induced magnetization $M_y(x)$ in metallic films with thicknesses of $d=10$ nm (solid line), $20$ nm (dashed line), and $30$ nm (dotted line).
    The red lines correspond to the magnetization distribution due to the binding mode $M_{y,-}(x)$,
    and the blue lines are the same but for the antibinding mode $M_{y,+}(x)$ (they almost overlap with each other).
    In this plot, the wavelength of the SPs is $k = \omega_p/c$.
    The magnetisation gradient becomes steeper as the film thickness decreases.
    As done for Fig. \ref{fig:dispersion_SP},
    we use the Drude parameter for gold ($\omega_p = 2.1 \times 10^{15}$) Hz, the permittivity of glass $\epsilon_i=2.25$, and use $\Phi_0 = 1$ for simplicity.
  }
  \label{fig:distribution_spin}
\end{figure}

\section{Electron-spin current pumped by surface plasmons on metallic film}
The spin diffusion equation \cite{valet1993theory} in the presence of the source current $\vec{j}_s^{\mathrm{sou}}$ is
\begin{equation}
  \left( \partial_t - D_s \nabla^2 + \frac{1}{\tau} \right) \delta \mu = \sigma_0^{-1} D_s \nabla \cdot \vec{j}_s^{\mathrm{sou}},
  \label{eq:spin_diffusion_equation}
\end{equation}
where the spin diffusion constant $D_s\equiv\lambda_s^2/\tau$,
with $\lambda_s$ being the spin diffusion length,
and $\tau$ the relaxation time.

We now study whether spin accumulation $\delta \mu$,
which drives spin currents,
is generated by the source term that stems from the SP-induced magnetization gradient:
\begin{equation}
  \vec{j}_s^{\mathrm{sou}} = -\frac{\hbar \sigma_0}{m}\nabla M_y.
  \label{eq:source}
\end{equation}
Here we focus on the stationary state of the system.
The stationary diffusion equation can be obtained by eliminating the time-derivative term and by substituting the expression for the source term Eq. (\ref{eq:source}) into Eq. (\ref{eq:spin_diffusion_equation}).
The result is
\begin{equation}
  \nabla^2 \delta \mu =\frac{\delta \mu}{\lambda_s^2} +\frac{\hbar}{m}\nabla^2 M_y.
  \label{eq:spin_diffusion_equation_stationary}
\end{equation}
The stationary solution to this differential equation is
\begin{equation}
  \delta \mu = \frac{\hbar}{m}\frac{(2 K_m \lambda_s)^2}{(2 K_m \lambda_s)^2 - 1}M_y.
  \label{eq:spin_acc}
\end{equation}
This is the spin accumulation induced by the SPs in the metallic film.
The resonance conditions are given by
\begin{equation}
  (2 K_m \lambda_s)^2 - 1 \rightarrow 0,
  \label{eq:resonance}
\end{equation}
which is determined both by the wavenumber of the SPs $k$ and by the spin diffusion length $\lambda_s$ of the metal.
In addition,
the resonance condition depends on the SP modes and on the thickness of the film because $K_m$ is a function of $\omega = \omega_\pm(k,d)$,
where $\omega_+$ and $\omega_-$ are different in general.

\par
\begin{figure}[tbhp]
  \centering
  \includegraphics[width=.9\linewidth]{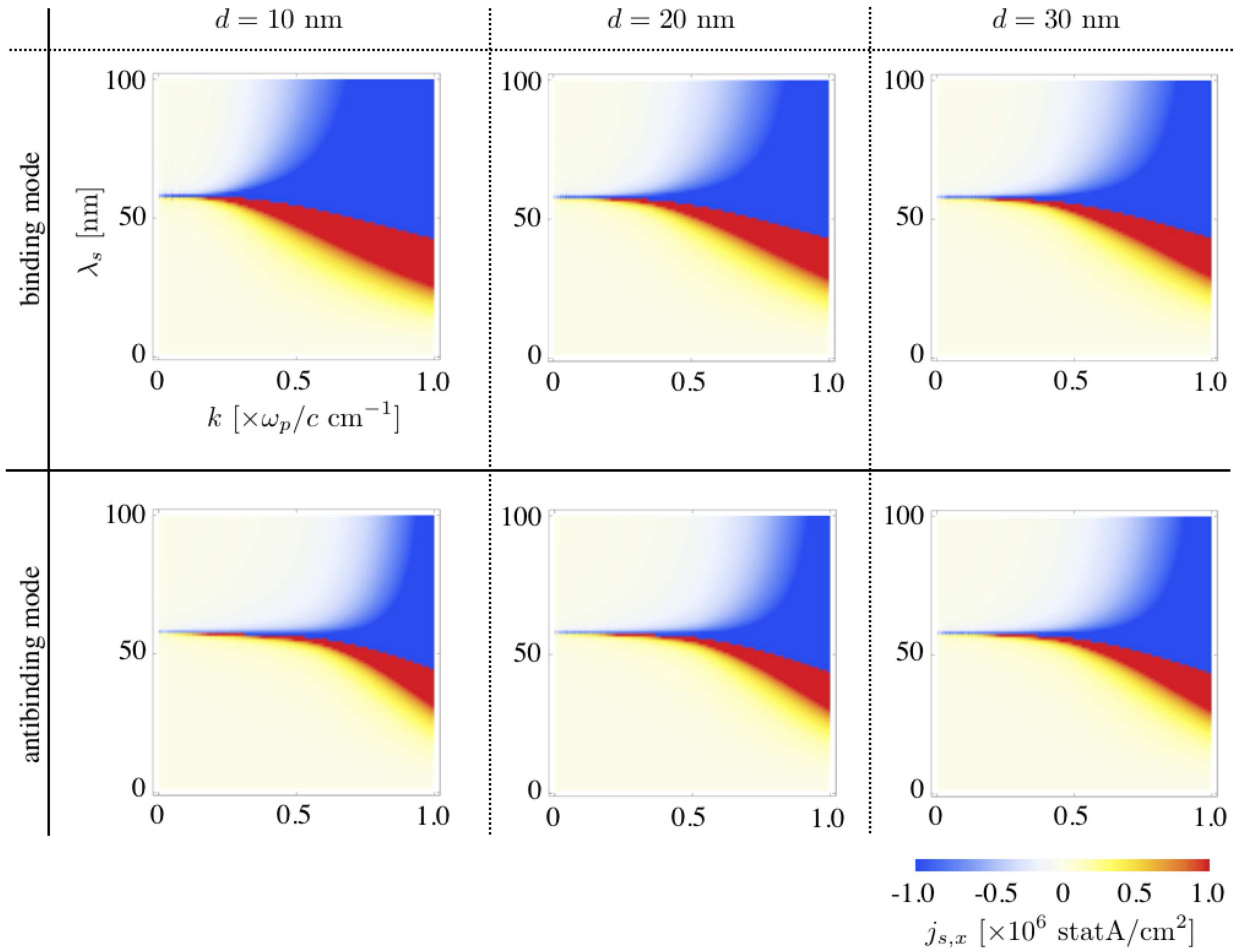}
  \caption{
    Diffusive spin current as a function of the wavenumber $k$ of the SPs and the spin diffusion length $\lambda_s$ of the metallic film for various film thicknesses ($d=10,\ 20,\ 30\ \mathrm{nm}$).
    For these plots, $x = d/2$, and $\omega_\mathrm{p}=2.1 \times 10^{15}\ \mathrm{Hz}$ and $\epsilon_i=2.25$ as in all previous figures and $\Phi_0 = 1$ for simplicity.
    Note that the plot range of the spin current goes from $-1.0 \times 10^6\ \mathrm{statA/cm}^2$ to $1.0 \times 10^6\ \mathrm{statA/cm}^2$ because the peak value of the current is so large.
    The spin currents are resonantly driven when the resonance conditions are met,
    which occurs when the sign of the spin current flips.
    A larger spin current is generated for thinner films.
  }
  \label{fig:js_diff_map}
\end{figure}

The diffusive spin current driven by the SP-induced spin accumulation Eq. (\ref{eq:spin_acc}) is
\begin{equation}
  \vec{j}_s = \frac{(2 K_m \lambda_s)^2}{(2 K_m \lambda_s)^2 - 1} \vec{j}_s^{\mathrm{sou}}.
  \label{eq:js}
\end{equation}
FIG. \ref{fig:js_diff_map} shows the diffusive spin current near the upper interface as a function of wavenumber and for various film thicknesses.
The resonance conditions are clearly observed when the sign of the spin currents is flipped.
This result implies that the direction of the spin current can be controlled by tuning the wavelength of the SPs and/or the spin diffusion length of the metal.

\par
Finally, we estimate the magnitude of the driven spin currents.
For both the binding and the antibinding modes,
the magnetization $\vec{M}$ is of the order of $10^{-3}\ \mathrm{G} \approx 10^{-6}\ \mathrm{T}$ when $\Phi_0 = 10^{-4}\ \mathrm{statV}$,
which is equivalent to the electric field of a laser with an intensity of $10\ \mathrm{mW}/\mathrm{cm^2}$ and a focal-spot size of $10^4\ \mathrm{\mu m^2}$.
A decay length for SPs on the order of $100\ \mathrm{nm}$ gives a magnetization gradient
\begin{equation*}
  \nabla M_y \sim \frac{1}{100 \times 10^{-9}} \times 10^{-6} = 10\ \mathrm{T/m}
\end{equation*}
The source term $\vec{j}_s^{\mathrm{sou}} \approx 10^5\ \mathrm{A/m^2}$.
Thus,
for the binding mode,
the resonance coefficient $\frac{(2 K_m \lambda_s)^2}{(2 K_m \lambda_s)^2 - 1} \sim 100$ at one of the peaks ($k = 83127.7\ \mathrm{cm^{-1}}$, $\lambda_s = 60\ \mathrm{nm}$) for the binding mode,
so the driven spin current $\vec{j}_s$ is of the order of $10^7\ \mathrm{A/m^2}$,
which is large enough to detect by the inverse spin Hall measurement \cite{ando2011inverse}.

\par
With decreasing film thickness,
the magnetization gradient becomes steeper because of the strong hybridization of the SPs,
thereby driveing a larger spin current at least within the low wavenumber region.
On the other hand, in the high wavenumber region, the SPs will be damped by several mechanisms such as interband transitions, electron-electron and electron-phonon interactions, and surface scattering \cite{voisin2001ultrafast, yuan2008landau, li2013landau, shahbazyan2016landau}.
In order to take these effects into consideration,
further analysis with quantum mechanical treatment is needed,
which we leave for future works.

\section{Conclusion}
In conclusion, we derived the eigenmodes of surface plasmons (SPs) in a thin-film system and calculated their spin angular momentum.
Solving the spin diffusion equation in the presence of the SPs reveals that spin current is generated by the SPs in the thin film.
The spin current is resonantly driven when a resonance condition is satisfied,
which occurs when the polarization of the driven spin current flips.
At one resonance condition (for the binding mode, $k = 83127.7\ \mathrm{cm^{-1}}$, $\lambda_s = 60\ \mathrm{nm})$,
the driven spin current is about $10^7\ \mathrm{A/m^2}$,
which can be detected by an inverse spin Hall measurement.
The SPs become more strongly hybridized in thinner films and thereby generate a steeper magnetisation gradient,
which increases the driven spin current.
These results should serve to connect the fields of plasmonics and spintronics.

\ack
  This work is partially Supported by the Priority Program of Chinese Academy of Sciences, Grant No. XDB28000000.

\section*{References}
\input{manuscript.bbl}

\bibliographystyle{iopart-num}

\end{document}

%% file: manuscript.bbl
\providecommand{\newblock}{}